\documentclass[conference,compsoc]{IEEEtran}
\ifCLASSOPTIONcompsoc
  \usepackage[nocompress]{cite}
\else
  \usepackage{cite}
\fi
\ifCLASSINFOpdf
  \usepackage[pdftex]{graphicx}
  \DeclareGraphicsExtensions{.pdf,.jpeg,.png}
\else
  \usepackage[dvips]{graphicx}
  \DeclareGraphicsExtensions{.eps}
\fi
\ifCLASSOPTIONcompsoc
  \usepackage[caption=false,font=footnotesize,labelfont=sf,textfont=sf]{subfig}
\else
  \usepackage[caption=false,font=footnotesize]{subfig}
\fi
\usepackage{url}

\usepackage{rotating}

\usepackage{arydshln}
\usepackage{framed}

\begin{document}
%
\title{ClouNS - A Cloud-native Application Reference Model for Enterprise Architects}

\author{\IEEEauthorblockN{Nane Kratzke}
\IEEEauthorblockA{L\"ubeck University of Applied Sciences\\
Center of Excellence CoSA\\
L\"ubeck, Germany \\
Email: nane.kratzke@fh-luebeck.de}\and
\IEEEauthorblockN{Ren\'e Peinl}
\IEEEauthorblockA{Hof University of Applied Sciences\\
Institute of Information Systems \\
Hof, Germany \\
Email: rene.peinl@iisys.de}
 }


%


\maketitle

\begin{abstract}
The capability to operate cloud-native applications can generate enormous business growth and value. But enterprise architects should be aware that cloud-native applications are vulnerable to vendor lock-in.
We investigated cloud-native application design principles, public cloud service providers, and industrial cloud standards. All results indicate that most cloud service categories seem to foster vendor lock-in situations which might be especially problematic for enterprise architectures. This might sound disillusioning at first. However, we present a reference model  for cloud-native applications that relies only on a  small subset of well standardized IaaS services. The reference model can be used for codifying cloud technologies. It can guide technology identification, classification, adoption, research and development processes for cloud-native application and for vendor lock-in aware enterprise architecture engineering methodologies.
\end{abstract}


%
\IEEEpeerreviewmaketitle

\section{Introduction}

\noindent Even for small companies, it is possible to generate enormous economical growth and business value by  providing cloud-native services or applications. E.g., Instagram proofed successfully that it was able to generate exponential business growth and value in a very short amount of years. At the time of being bought by Facebook for 1 billion USD, Instagram had only a headcount of about 20 persons, was only two years old, but operated already a world wide accessible and scalable social network for image sharing (hosted by Amazon Web Services, AWS) without owning any data center or any noteworthy IT assets. On the other hand, if AWS had to shutdown their services at that time due to an insolvency or something alike, it is likely that Instagram had been carried away by such an event as well. The reader should consider, that it took years for Instagram's cloud engineers to transfer their services from the AWS cloud infrastructure completely into Facebook's data centers. And this transfer was accompanied by several noteworthy service outages. 

Enterprise architects should be aware of such kind of vendor lock-in situations and should try to avoid them by design.
According to Pahl et al.  cloud computing must implement  universal  strategies  regarding  standards,  interoperability  and  portability to reduce vendor lock-in situations.  Open  standards  are  of  critical  importance and need to be embedded into interoperability solutions \cite{PZF2013}. However, standardization processes are slow. We found that 80\% of provided public cloud services are not even considered in cloud standards like \textit{CIMI} \cite{CIMI} or \textit{OCCI} \cite{OCCI-CORE, OCCI-IS}. This might sound disillusioning. But it might be the key to handle vendor lock-in pragmatically, if we simply accept this as a fact. At last it means, that there is a small part of standardized cloud computing. We propose to use this "isle of felicity" as the foundation of a portable cloud runtime environment. Satzger et al. developed already such a vision for a vendor lock-in free cloud. "\textit{A \textbf{meta cloud} [...] incorporates design time and runtime components. This meta cloud would abstract away from existing offerings’ technical incompatibilities, thus mitigating vendor lock-in. It [...] supports an application’s initial deployment and runtime migration.}" \cite{SHI+2013}. They further postulated that "\textit{most of the basic technologies necessary to realize the \textit{meta cloud} already exist, yet lack integration}". Our presented ClouNS reference model is in accordance with this \textit{meta cloud} concept but it is much more concrete than visionary. It integrates existing container solutions (like \textit{Docker}), existing cloud infrastructure bridging environments (scalable and container based cloud runtime environments like \textit{Mesos/Marathon} or \textit{Kubernetes}), existing microservice architecture approaches and existing cloud-native application patterns into one integrated concept which considers vendor lock-in awareness in each conceptual layer. All these identified building blocks of cloud-native applications can be valuable components of cloud-based and vendor lock-in aware enterprise architectures.

\section{Outline and research methodology}

\begin{figure}[tb]
	\begin{center}
		\includegraphics[width=\columnwidth]{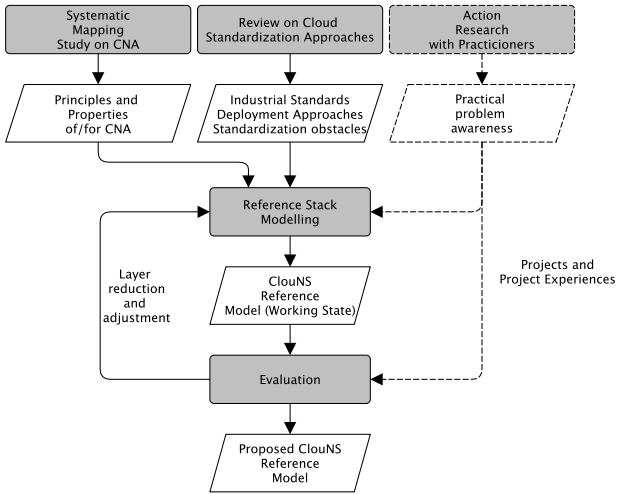}
	\end{center}
	\caption{Research methodology}
	\label{fig:methodology}
\end{figure}

\noindent Our research methodology is shown in Figure \ref{fig:methodology}. We performed a systematic mapping study on cloud-native applications to get a better understanding what cloud-native applications (CNA) exactly are. Basic insights are summarized in Section \ref{sec:cloud-native-app}. Because cloud-native applications are very vulnerable to vendor lock-in, we performed additionally a review on cloud standardization approaches. Section \ref{sec:approaches-to-avoid-vendor-lockin} summarizes how vendor lock-in emerges in cloud computing. Both reviewing steps have been accompanied by action research in concrete projects or by having cloud-native application engineering approaches by practitioners under research surveillance. Our resulting and proposed reference model ClouNS (cloud-native application stack) is presented in Section \ref{sec:clouns} and relies only on a small subset of standardized IaaS services to avoid vendor lock-in. We evaluated this reference model in  Section \ref{sec:evaluation}  using a concrete project from our action research activities and show that the presented ClouNS reference model covers more aspects of cloud-native applications than any other current cloud standard.

\section{Cloud-native applications}
\label{sec:cloud-native-app}
\noindent It is common sense that cloud-native applications are developed intentionally for the cloud. However, there is no common definition that explains what that exactly means. Our literature review\footnote{Currently prepared for a separate publication.} on cloud-native applications turned out, that there is a common but unconscious understanding across all analyzed papers. Cloud-native applications (CNA) should be build according to corresponding CNA principles (operated on automation platforms providing softwarization of infrastructure and network, having migration and interoperability aspects in mind). These principles enable to build CNA architectures with specific and desirable CNA properties (horizontal scalablity, elasticity, resiliency, isolated states, to name some of the most mentioned properties). Accompanying CNA methodologies are often pattern based. The following insights are the most valuables from authors' point of view.

Fehling et al. propose that a cloud-native application should be IDEAL. It should have an \textbf{\underline{i}solated state}, is \textbf{\underline{d}istributed} in its nature, is \textbf{\underline{e}lastic} in a horizontal scaling way, operated via an \textbf{\underline{a}utomated management} system and its components should be \textbf{\underline{l}oosely coupled} \cite{FLR+2014}. According to Stine there are common motivations for cloud-native application architectures \cite{Stine2015} like 
to deliver software-based solutions more quickly (\textbf{speed}), 
in a more fault isolating, fault tolerating, and automatic recovering way (\textbf{safety}),
to enable horizontal (instead of vertical) application scaling (\textbf{scale}),
and finally to handle a huge diversity of (mobile) platforms and legacy systems (\textbf{client diversity}).

These common motivations are adressed by several application architecture and infrastructure approaches \cite{BHJ2015}: \textbf{Microservices} represent the decomposition of monolithic (business) systems into independently deployable services that do "one thing well" \cite{Newman2015,namiot2014}. 
The main mode of interaction between services in a cloud-native application architecture is via published and versioned APIs (\textbf{API-based collaboration}). These APIs often follow the HTTP REST-style with JSON serialization, but other protocols and serialization formats can be used as well.
Single deployment units of the architecture are designed and interconnected according to a \textbf{collection of cloud-focused patterns} like the \textit{twelve-factor app} collection \cite{Wiggins2012}, the \textit{circuit breaker} pattern \cite{Fowler2014} or cloud computing patterns \cite{FLR+2014}.
And finally, \textbf{self-service agile infrastructure platforms} are used to deploy and operate these microservices via self-contained deployment units (containers). These platforms provide additional operational capabilities on top of IaaS infrastructures like automated and on-demand scaling of application instances, application health management, dynamic routing and load balancing as well ass aggregation of logs and metrics.

These aspects lead us to the following definition:
\begin{framed}
	\noindent A \textbf{cloud-native application} is a distributed, elastic and horizontal scalable system composed of (micro)services which isolates state in a minimum of stateful components. The application and each self-contained deployment unit of that application is designed according to cloud-focused design patterns and operated on a self-service elastic platform.
\end{framed}


\section{Approaches dealing with vendor lock-in}
\label{sec:approaches-to-avoid-vendor-lockin}
\noindent Vendor lock-in is defined to make a customer dependent on a vendor for products and services, unable to use another vendor without substantial switching costs \cite{lockin2006,Arthur1989}. This section will explain how vendor lock-in emerges and how it is typically addressed looking at relevant cloud computing use cases identified by \cite{di2015class}.

\smallskip
\noindent \textbf{Vendor lock-in in cloud computing.} 
\noindent A commodity describes a class of demanded goods (in case of IaaS computing, storage and networking services) without substantial qualitative differentiation. The market treats instances of the goods as (nearly) equivalent. Or more simple: a virtual server provided by Google Compute Engine (GCE) can easily being replaced by a server provided by Amazon Web Services (AWS, i.e. system portability in NIST terms), as described in use case CCUC 3 \cite{di2015class} and one can even find very similar instances regarding resources and performance \cite{KQ2015}.  Therefore, IaaS provides a commodity service.

On the other hand, switching from one SaaS service provider to another one (CCUC 1, \cite{di2015class}) or changing the middleware (PaaS, CCUC 2) is in most cases not as easy as buying coffee from a different supplier. While CCUC 1 is about finding a different provider with the same SaaS offering (e.g., hosted SharePoint) and migrating configuration and data (data portability in NIST terms), this paper concentrates on CCUC 2 and the respective tasks of application architects to build their cloud-native application on services of one provider, that are easily replacable by equivalent services of other providers (application portability in NIST terms \cite{di2015class}).
The core components of these distributed systems like virtualized server instances and basic networking and storage can be deployed using commodity services. However, further services needed to integrate these virtualized resources in an elastic and scalable manner are often not considered in standards. Services like load balancing, auto scaling or message queuing systems are needed to design an elastic and scalable cloud-native system on any cloud service infrastructure. But especially these services are not considered consequently by current cloud standards (see Table \ref{tab:cloud_services} and \cite{Kra2014a}). All cloud service providers try to stimulate cloud customers to use these non-commodity convenience services in order to bind them to their infrastructure.
So, the following strategies of top-down standardization, open source bottom-up solutions and formalized deployment approaches can be identified to overcome this kind of vendor lock-in. For a more detailed discussion we refer to \cite{OSF2014}.

\smallskip
\noindent \textbf{Industrial Top-Down Standardization Approaches.} We cover only the following few approved service interoperability standards \cite{HFST2011}. For a more complete survey on cloud computing standards we refer to \cite{PZF2013}.

\textit{Cloud Infrastructure Management Interface (\textbf{CIMI}) specification} \cite{CIMI} describes the model and protocol for management interactions between a cloud Infrastructure as a Service (IaaS) provider and the consumers. The basic resources of IaaS (machines, storage, and networks) are modeled with the goal of providing consumer management access to an implementation of IaaS and facilitating portability between cloud implementations that support the specification. CIMI is accompanied by the \textit{Open Virtualization Format }(OVF). OVF describes an \textit{"open, secure, portable, efficient and extensible format for the packaging and distribution of software to be run in virtual machines"} \cite{OVF}.

The \textit{Open Cloud Computing Interface (\textbf{OCCI})} \cite{OCCI-CORE,OCCI-IS} is a RESTful Protocol and API. OCCI was originally initiated to create a remote management API for IaaS services,
allowing for the development of interoperable tools for common tasks including deployment, autonomic scaling
and monitoring. It has since evolved into a flexible API with a focus on interoperability and offering
a high degree of extensibility.

The \textit{Cloud Data Management Interface (\textbf{CDMI})} \cite{CDMI} is used to create, retrieve,
update, and delete objects in a cloud via a RESTful API. The CDMI reference model focuses Storage as as Service and covers block storage, file system storage, object storage, table-based storage and storage for large scale queueing systems. Due to its storage focus it is more limited than CIMI and OCCI. 

There are further accompanying drafts, interoperability standards, guides and working groups like \cite{SIIF,CPIP} having less concrete working results than the already mentioned standards. So, we did not list them in Section \ref{sec:evaluation} (Table \ref{tab:cloud_services}) for our evaluation. 

\smallskip
\noindent \textbf{Open-Source Bottom Up Approaches.} Open source bottom up approaches try to harmonize the API of cloud providers with an own abstraction (e.g., Deltacloud, fog.io, jClouds, Apache Libcloud, see \cite{dimartino2015}). They try to do the same like the industrial standardization approach but apply a bottom up strategy. These kind of libraries provide (limited) access to 15 - 30 cloud service providers and support:
Cloud Servers and Block Storage, Object Storage and CDN, Load Balancing and DNS as a Service, Container Services.
In fact, that is very similar compared to the coverage of industrial standards.

\smallskip
\noindent \textbf{Formalized Deployment Approaches}. Several research approaches tried not to focus the cloud infrastructure as object of standardization but to standardize the deployment of complex systems. This is often done by defining descriptive (often XML-based) deployment languages \cite{JD2011,ACA+2011,LDK+2011,MGGD2012}. However, these kind of languages are often limited to static (non elastic) deployments. To handle elasticity these approaches need something like a runtime environment responsible for auto-scaling and load balancing.
All of these research directions were valuable and contributed to standardization approaches like TOSCA \cite{TOSCA}. More recent research approaches are SALSA focusing dynamic configuration of cloud services \cite{LTC+2014}, SPEEDL focusing an event-based language to define the scaling behavior of cloud applications \cite{zlsd2015} or SYBL as an extensible language for controlling elasticity in cloud applications \cite{CMT+2013}.

\smallskip
\noindent \textbf{Summary.} 
\label{sec:critical-view}
 The release cycles of new cloud services over the last 10 years show that standardization processes are far slower than cloud service development and deployment processes of major public cloud providers. This phenomenon is not new in information technology and often fosters establishing so called de facto standards like MS-DOS and IBM PCs in the 1980's, Windows/MS-Office or PDF in the 1990's, Java Virtual Machine in the 1995's, Android and iOS in 2010's, just to name a few.

All reviewed cloud standards and open-source bottom up approaches concentrate on a very small but basic subset of popular cloud services: compute nodes, storage (file, block, object), networking, logging and (static) system deployments. Also deployment approaches are defined against this infrastructure level of abstraction. These kind of services are often subsumed as IaaS and build the foundation of cloud services and therefore cloud-native applications. All other service categories might foster vendor lock-in situations if they are not standardized and can not be provided in a form to be deployed on any \textit{"portable cloud runtime environment"}. This all might sound disillusioning. But in consequence, the basic idea of a cloud-native application stack used for enterprise architectures should be to use only this small subset of well standardized IaaS services as founding building blocks (see Section \ref{sec:clouns}).

\section{The cloud-native application stack}
\label{sec:clouns}


\noindent The Open Systems Interconnection model (OSI model) 
is a well accepted conceptual model that characterizes and standardizes the communication functions of a communication system without regard to their internal structure and technology \cite{ISO-7498}. Its goal is the interoperability of diverse communication systems with standard protocols. The model partitions a communication system into abstraction layers. Although the OSI model can not be applied directly to cloud service interoperability, the underlying principles of the layered architecture can be applied to a cloud stack as well:

\begin{small}
\begin{enumerate}
\item Create so many layers as necessary but not more.
\item Create a boundary at a point where the description of services can be small and interactions across boundaries are minimized.
\item Create separate layers to handle functions that are different in the process performed or the technology involved.
\item Collect similar functions into the same layer.
\item Create layers of localized functions. A layer must be totally redesignable to take advantage of new advances in architectural, hardware or software technology without changing services expected from and provided to the adjacent layers.
\item Allow changes of functions or protocols to be made within a layer without affecting other layers.
\item Create for each layer, boundaries with its upper and lower layer only.
\end{enumerate}
\end{small}

\noindent The same principles were successful in the past to structure a vast amount of fast evolving network technologies, tools and standards. Why should this not be working for the cloud computing domain which is characterized by a vast amount of fast evolving cloud technologies, tools and standards? Our reference model is intentionally designed to be enterprise architecture framework agnostic. Due to the great amount of EA frameworks, it should be composable with existing approaches like TOGAF, FEA, NAF, etc. \cite{WBM+2010}. However, we will show exemplary (if appropriate) how our approach relates to EA frameworks like TOGAF for instance.

\begin{table*}
\centering
\footnotesize
\renewcommand{\arraystretch}{1.1}
\caption{Layers of a cloud-native stack (ClouNS reference model)}
\label{tab:clouns_layers}
\begin{tabular}{ l l l l l l }
\hline
\textbf{Viewpoint}       & \textbf{Layer} & \textbf{(Sub)layer Name}                        & \textbf{Main Purpose}                               & \textbf{Examples}  & \textbf{Example Standards} \\
\hline
SaaS & (6) Application     & Application                 & Providing End User Functionality    &                                                &  \\
\hline
PaaS        &  (5) Service     & Functional Services   &  Providing Functional Services & RabbitMQ, Hadoop                      &  AMQP \\
               &                       & All Purpose Services   &  Providing Distrib. Sys. Patterns & SpringCloud                      &   \\
                 &         & Storage Services   &  Providing Storage Services & S3, Swift, RADOS                      & CDMI \\
                  &         &                                  & Providing Database Services & RDS & SQL \\
        
\hline
CaaS        & (4) Cluster      & Container Orchestrator &  Providing Continual Elasticity      & Kubernetes, Marathon                & TOSCA \\
                &       & Overlay Network        & Bridging IaaS Networks           & Flannel, Weave, Calico           &  \\
                  &       & Cluster Scheduler        & Scaling Cluster              & Swarm, Mesos, ECS          &  \\
                  &       &                                 &  Bridging IaaS Infrastructures       &           &  \\
                  &       & Clustered Storage      & Providing Scalable Storage           & Gluster FS, Ceph          & NFS, CIFS, WebDav \\                  
\hline
IaaS          & (3) Container Host      & Container                   & Executing Deployment Units     & Docker, Rkt                         & OCI \\
                &                                  & Operating System       & Operating Hosts  & Linux, OS X, Windows       & POSIX \\
                   & (2) Virtual Host      & Virtual Infrastructure   & Providing Virtual Machines             & EC2, GCE, OS, ESX  & OVF, OCCI, CIMI \\
\hdashline
                   & (1) Physical Host      & Physical Infrastructure & Providing Host Machines                & Bare Metal Machines       & \\
\end{tabular}
\end{table*}

\begin{figure*}[tb]
	\centering
	\includegraphics[width=0.995\textwidth]{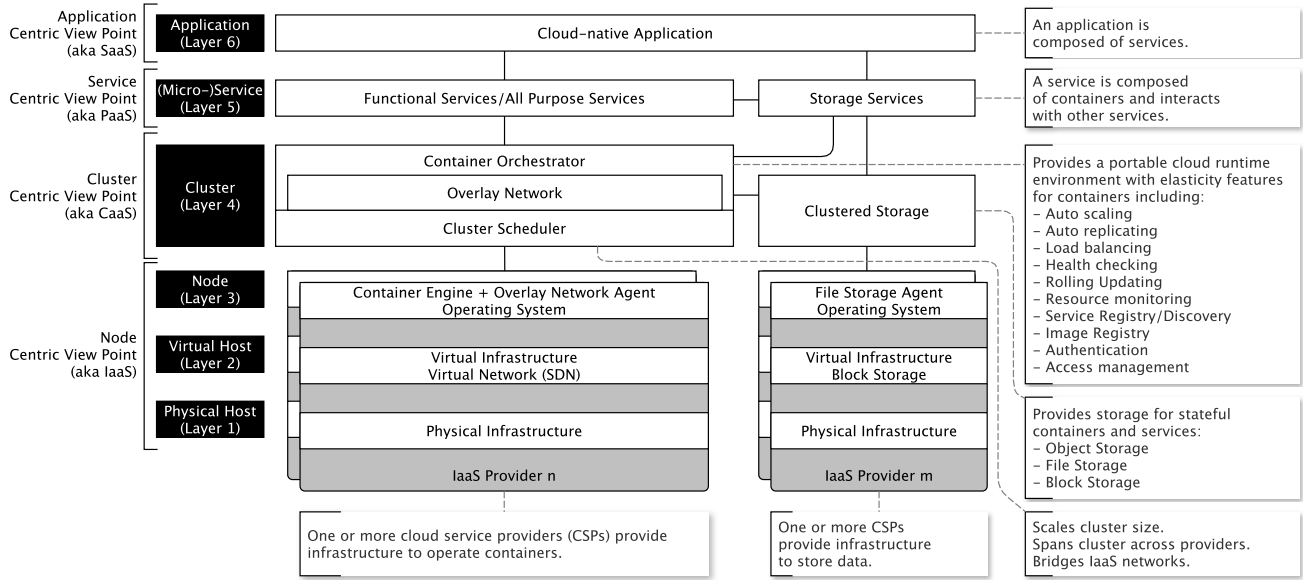}
	\caption{The Anatomy of the Cloud-Native Stack (ClouNS)}
	\label{fig:cloud_native_stack}
\end{figure*}

To have multiple view points on the same object was made popular by the Zachman Framework \cite{Zachman2008} and has been adapted successfully by other architecture frameworks and methodolgies.  Obviously the same is true for cloud computing. We can look from a service model point of view (IaaS, PaaS, SaaS), a deployment point of view (private, public, hybrid, community cloud) on cloud computing as it is done by \cite{MG2011}. Or we can look from an actor point of view (provider, consumer, auditor, broker, carrier) or a functional point of view (service deployment, service orchestration, service management, security, privacy) as it is done by \cite{BML+2011}. Points of view are particular useful to split problems into concise parts. However, the above mentioned view points are useful from service provider point of view but not from cloud-native application engineering point of view. From an engineering point of view it would be more useful to have views on technology levels involved and applied in cloud-native application engineering as it is often done by practitioner models. However, these practitioner models have been only documented in some blog posts\footnote{Practicioner Blog Posts:

  Jason Lavigne, "\textit{Don't let aPaaS you by - What is aPaaS and why Microsoft is excited about it}", see \url{https://atjasonunderscorelavigne.wordpress.com/2014/01/27/dont-let-apaas-you-by/} (last access 1th May 2016)
   
  Johann den Haan, "\textit{Categorizing and Comparing the Cloud Landscape}", see \url{http://www.theenterprisearchitect.eu/blog/categorize-compare-cloud-vendors/} (last access 1th May 2016)
} and do not expand into any academic papers as far as the authors know. Therefore, we propose the following four basic view points for ClouNS (see Table \ref{tab:clouns_layers} and Figure \ref{fig:cloud_native_stack}) which can be aligned to TOGAF concepts like business, information systems, and technology architecture:

\smallskip
\noindent \textbf{(1) Node centric view point (aka IaaS).} This is a view point being familiar for engineers who are developing classical client server architectures. The server parts are often deployed onto a single compute node (a server). This is how IaaS is understood most of the times. IaaS deals with deployment of isolated compute nodes for a cloud consumer. It is up to the cloud consumer what it is done with these isolated nodes (even if there are provisioned hundreds of them). EA frameworks like TOGAF covering this view point often with a technology architecture.

\smallskip
\noindent \textbf{(2) Cluster centric view point (aka CaaS).} This is a view point being familiar for engineers who are dealing with horizontal scalability across nodes. Clusters are a concept to handle many nodes as one logical compute node (a cluster). Such kind of technologies are often the technological backbone for PaaS solutions and portable cloud runtime environments because they are hiding complexity (of hundreds or thousands of single nodes) in an appropriate way. We call these platforms portable because they can be deployed on any IaaS infrastructure. Recently the term CaaS (Container as a Service) is more and more used to describe such solutions if they are using container technologies under the hood. Furthermore, we have to consider scalable cluster storage solutions necessary for stateful services like Database as a Service approaches. According to TOGAF this view point is reflected mainly in the technology architecture as well. Additionally, CaaS realizes the foundation to define services and applications without reference to particular cloud services or cloud infrastructures and therefore provides the basis to avoid vendor lock-in.

\smallskip
\noindent \textbf{(3) Service centric view point (aka PaaS).} This is a view point familiar for application engineers dealing with Web services in service-oriented architectures (SOA). Of course, (micro)services have to be deployed on and operated by single nodes. However, all necessary and complex orchestration of these single nodes is delegated to a cluster (cloud runtime environment) providing a platform as a service (PaaS). According to TOGAF this layer is covered mainly in the information systems architecture. Due to the fact that microservices are aligned to business capabilities, this fits very well with the TOGAF intent, that an application architecture should mainly identify \textit{"logical groups of capabilities that manage data and support business functions"}.

\smallskip
\noindent \textbf{(4) Application centric view point (aka SaaS).} This is a view point being familiar for end-users of cloud services (or cloud-native applications). These cloud services are composed of smaller cloud services being operated on clusters formed of single compute and storage nodes. The cloud provision model SaaS falls into this category. However, most of the times -- and on this level -- cloud users are not interested in underlying technological details and layers, they are thinking in business processes, business services, and business functions. Therefore, this layer is covered mainly in the TOGAF information systems (application) architecture. Due to the fact, that it is most of the times business use-case oriented it has much clearer relationships to business goals, functions, services, and processes identified in a TOGAF business architecture.

\subsection{Node-centric view point aka IaaS }

\noindent A node centric view on cloud computing comprises several layers to handle the physical infrastructure, the virtual infrastructure (main outcome of IaaS services), the operating systems layer provided inside virtual machines (VMs) and the currently popular container layer providing self-contained deployment units (so called containers). Although currently it can be considered good practice to operate containers on VMs, ongoing developments to increase security and encapsulation of containers make it probable that the virtualization layer will loose importance in the future as containers will directly run on the OS atop the physical hardware (like Google is said to do already). 

\smallskip
\noindent \textbf{Layer 1: Physical Host.} This layer is the cloud computing analogy for the physical media  of the OSI model. Due to the fact that this is normally hidden and only of direct access to a cloud provider, we will not cover this layer in detail. 

\smallskip
\noindent \textbf{Layer 2: Virtual Host.} This layer is the domain of IaaS solutions. It is mainly covered by cloud standards like CIMI, OCCI, OVF. So, Layer 2 is one of the best industry standardized layers in the presented reference model. Typical public or private cloud services are the \textit{EC2} service by \textit{AWS}, \textit{GCE} by \textit{Google}, \textit{Virtual Machines} service by \textit{Azure} or \textit{Nova} by \textit{OpenStack}.  This layer provides\textbf{ virtual machines}, which are connected by a \textbf{network} of the cloud provider. The machines can be configured to be accessible from public internet via configurable IP-source and port ranges. 
The machines can be assigned \textbf{disk storage} provided by the cloud infrastructure.

\smallskip
\noindent \textbf{Layer 3: Container Host.} In most cases, standard operating systems are installed on virtual machines to operate the node and to install and configure additional arbitrary software packages. 
The most prominent operating system related standard might be the  POSIX standard family \cite{POSIX}. Typical operating systems used in cloud computing are often Linux/Unix systems, but arbitrary operating systems are possible. There are some Linux based OS like \textit{CoreOS} or \textit{Atomic} especially configured for providing only \textbf{container runtime environments} (see \cite{PH2015} for more), but machines on Layer 3 can be used as all purpose nodes.
However, there is a trend to use container technologies to simplify deploying and minimize incompatibilities of different applications running on the same host.
Container solutions, notably Docker \cite{PH2015}, provide a \textbf{self-contained} \textbf{standard runtime}, \textbf{image format}, and \textbf{build system} for operating system containers deployable to any Layer 3 machine. The open container initiative (OCI, \cite{OCI}) is an open governance structure with the purpose of creating open industry standards around container formats and runtime environments focusing this layer.

\subsection{Cluster view point  aka CaaS }

\noindent \textbf{Layer 4: Cluster.} A cluster centric view on cloud computing comprises several sublayers to integrate multiple Layer 3 container hosts into one single and higher level logical container cluster (Container as a Service, CaaS \cite{caas}). These technologies are used as \textbf{self-service agile infrastructure platforms} \cite{Stine2015} and they provide a \textbf{portable cloud runtime environment} for cloud-native applications. These platforms are complex, so we have to distinguish different components of such a container cluster. \textit{Apache Mesos} \cite{mesos} is a prominent type representative here. It has been successfully operated for years by companies like Twitter or Netflix to consolidate hundreds of thousands of compute nodes. More recent approaches are \textit{Docker Swarm} and Google's \textit{Kubernetes}, the open-source successor of Google's internal \textit{Borg} system (see \cite{kubernetes} and \cite{PH2015} for an overview). There are a lot of variants of deploying these tools. It is possible to deploy \textit{Kubernetes} on \textit{Mesos} or on \textit{Swarm}. By default, \textit{Kubernetes} uses \textit{CoreOS/fleet} as a cluster scheduler. And all technologies providing a clustering technology around container. But what is the difference?

\textbf{(1) Scheduler.} From our point of view there are three main benefits of cluster schedulers like \textit{Docker Swarm} or \textit{Apache Mesos}, starting with the \textbf{integration of single nodes (container hosts) into one logical cluster (1st benefit)}. This integration can be done within an IaaS solution (for example only within \textit{AWS}) and is mainly done for complexity management reasons. However, it is possible to deploy such cluster schedulers across public and private cloud infrastructures (for example deploying some Layer 3 container hosts of the cluster to \textit{AWS}, some to \textit{GCE} and some to an on-premise \textit{OpenStack} infrastructure). Even if the \textbf{container cluster is deployed across different cloud service providers (2nd benefit)} it can be accessed as logical single cluster, which is of course a great benefit from a \textbf{vendor lock-in avoiding (3rd benefit)} point of view. Our own research deals with aspects on this layer \cite{QK2016a}. 

\textbf{(2) Overlay Network.}
Even if only a single cloud service provider is involved, you need an easy way to communicate between containers on different hosts. That is why \textbf{overlay networks} like \textit{Weave} \cite{weave}, \textit{Calico} \cite{calico} or \textit{flannel} gained substantial interest. \textit{Weave} and \textit{Calico} 
can be integrated easily into \textit{Docker Swarm} (replacing the default \textit{libnetwork} overlay solution). \textit{Flannel} \cite{flannel} is well integrated with the \textit{CoreOS/fleet} cluster scheduler. Industrial standardization is quite rare on this layer. The authors are not aware of any  cloud-specific applicable standard (except general networking standards).

\textbf{(3) Orchestrator.} While cluster schedulers focus on the infrastructure unification, container orchestrators focus the payload which is being deployed to the cluster. \textit{Kubernetes} \cite{kubernetes} is a good example that container orchestrators can use basic deployment capabilities of cluster schedulers to deploy applications in a distributed and containerized form. There are several approaches how to operate Kubernetes on top of \textit{CoreOS/fleet}, on top of \textit{Docker Swarm} or on \textit{Mesos}. Docker itself calls this \textit{Swarm Frontends} and provides solutions to deploy \textit{Kubernetes} and \textit{Mesos/Marathon} on top of \textit{Docker Swarm}.

Additionally, container orchestrators provide \textbf{deployment unit-centric orchestrating}. Deployment units can be single containers or \textit{pods}, which is a container with all tightly coupled containers like associated storage containers in case of \textit{Kubernetes}. The orchestrator's benefit is, that it provides additional concepts to pure deploying of containers, like \textbf{high availability} and \textbf{elasticity} features (see Figure \ref{fig:cloud_native_stack} and \cite{PHP2016} for more details on tools involved). These features build the foundation to run elastic and distributed services on top of such clusters (or better portable cloud runtime environments). 


\textbf{(4) Clustered Storage.}
Most of the storage clusters are not part of the container cluster but container clusters are connected to \textbf{scalable storage clusters} in order to handle stateful services. File storage could be realized as a service on Layer 5. However, that might have some disadvantages. Storage cluster solutions like \textit{Ceph}, \textit{Gluster} or \textit{Torus} need very tight integration with the operating system (Layer 3). So, normally they do not provide their components in a containerized form. Furthermore, Layer 4 orchestrator's already need mountable file systems on Layer 4 to provide concepts like transferable volumes for containers. It would violate layer design principles if a Layer 4 orchestrator would use a Layer 5 service. It is more than likely that this would end in a "bootstraping paradox". Both aspects led to the decision to handle clustered storage as specialized storage clusters on Layer 4. Typically the provided storage is mountable by container hosts via POSIX standards like NFS or alike. This approach makes it easy to integrate the reference model with EA frameworks like TOGAF. TOGAF defines a data architecture. The clustered storage component is the realizing technological component for such a TOGAF data architecture.

\subsection{Service/application view point aka (P/S)aaS}

\noindent A service centric view on cloud computing focuses non-localizable services. A service can be composed of a multitude of single deployment units (containers/\textit{pods}) which are deployed against and operated by a Layer 4 cluster. To get the maximum benefit regarding safety and scalability, these deployment units should be following cloud-focused software design patterns (e.g. \textit{12-factor app} or \textit{circuit breaker} pattern). However, the cluster is responsible to scale and balance all service related deployment units to optimize service level quality. The presented model does not make any assumptions about what kind of services are operated. Most of the non-standardized services identified in Table \ref{tab:cloud_services} are services on this Layer 5. Layer 5 services are optimized for API-based collaboration (service to service communication). Layer 6 applications focusing the end user and are composed of Layer 5 services.

\smallskip
\noindent \textbf{Layer 5: Services.} In principal, arbitrary kind of services can be operated. However, some main categories of services and their accompanying frameworks can be categorized.

\textbf{(1) Functional Services} provide special functional capabilities as a service with API-based collaboration access. There exist plenty of analytical (e.g. \textit{Hadoop}/\textit{MapReduce}), prognostic (e.g. machine learning frameworks like \textit{TensorFlow}), messaging (e.g. \textit{RabbitMQ}) and a lot of further frameworks. Everything as a Service (or XaaS) is an often heard term fitting very well with Layer 5. There exist single standards like AMQP for messaging on this layer. However, regarding all currently existing and provided public cloud services we have to conclude that standardization on this level seems very incomplete (see Table \ref{tab:cloud_services}). 

\textbf{(2) All Purpose Services} (like \textit{SpringCloud}) are used to simplify application and service development by providing proven cloud-focused development and scalable interaction patterns. Most of the functional or all purposes services on Layer 5 are not standardized at all (see Table \ref{tab:cloud_services}).

\textbf{(3) Storage Services} often provide a REST-based API to store data. So, they build a perfect fit for API-based collaboration. Example frameworks are \textit{Swift} (\textit{OpenStack}), \textit{S3} (\textit{AWS}), \textit{RADOS} (\textit{Ceph}). These kind of services are often covered by the \textit{CDMI} standard. And of course, these services are aligned to data architectures requested by EA frameworks like TOGAF.

\smallskip
\noindent \textbf{Layer 6: Application.} An application centric view on cloud computing focuses the end user. Every Layer 5 service could be a Layer 6 application as well, if the service is accessible via a human machine interface (often this is realized via a web interface or a specialized mobile application). It is often the case, that Layer 5 services provide a Layer 6 human machine interface as well (at least for some administration purposes). Layer 6 applications are composed of Layer 5 services. Typical Layer 6 application examples are \textit{GoogleDocs}, \textit{AWS WorkSpaces}, or every other SaaS representative.

\section{Evaluating the reference model}
\label{sec:evaluation}

\begin{table*}
	
	\scriptsize
	\setlength{\tabcolsep}{1mm}
	\renewcommand{\arraystretch}{1.025}
	\caption{Popular cloud infrastructures/ecosystems and their considering by cloud standards.\newline(Most services are not considered by cloud standards but addressed by proposed ClouNS reference model.)}
	\label{tab:cloud_services}
	\begin{tabular}{ l l l l l | c c c c c c | c}
		
		\textbf{Service Category} & \textbf{Google}                               & \textbf{Azure} & \textbf{AWS} & \textbf{OpenStack} & \begin{sideways} OCCI\end{sideways} & \begin{sideways}CIMI\end{sideways} & \begin{sideways}CDMI\end{sideways} & \begin{sideways}OVF\end{sideways} & \begin{sideways}OCI\end{sideways}  & \begin{sideways}TOSCA\end{sideways} & \begin{sideways}ClouNS\end{sideways} \\
		\hline
		Compute             & Compute Engine  & Virtual Machines               & EC2 & Nova & C & M & & M & & & 2,3 \\
		& Custom Machine Types             &                                        &               & Glance & C & M& & & & & 2 \\
		& App Engine                              & Cloud Services                   & Elastic Beanstalk &  & & & & & & & 5 \\
		&                                               & RemoteApp  & MobileHub + AppStream &  & & & & & & & 5 \\
		
		\hline                           
		Container            & Container Engine                & Container Service & ECS & Magnum & & & & & \textcircled{?} & & 3, 4 \\
		& Container Registry              &   & EC2 Container Registry  &  & & & & & \textcircled{?} & & 4 \\
		\hline
		Storage               & Cloud Storage                    &                             & S3                        & Swift & & & St & & & & 5 \\
		&                                          & Storage (Disks)     & EBS                        & Cinder & St & V & St & & & & 2  \\
		&                                          & Storage (Files)      & Elastic File System & Manila & St &   & St & & & & 4 \\
		& Nearline                             & Backup                 & Glacier                   &  & & & & & & & 5 \\
		&                                          & StorSimple            &                                &  & & & & & & & 5 \\
		&                                          &                             & Storage Gateway      &  & & & & & & & 1 \\
		
		\hline
		Networking          & Cloud Networking               & Virtual Network & VPN & Neutron & N & N & & & & & 2 \\
		&                                          & Express Route   & Direct Connect &  & &  & & & & & 1 \\
		&                                          & Traffic Manager &   &  & &  & & & & & 4 \\
		&                                          & Load Balancer   & ELB &  & &  & & & & & 4 \\     
		&                                          & Azure DNS        & Route 53 & Designate & & & & & & & 4 \\     
		&                                          & Media Services  & Elastic Transcoder &  & &  & & & & & 5 \\     
		&                                          & CDN                 & Cloud Front & &  & & & & & & 5 \\                                 
		& Cloud Endpoints                  & Application Gateway & API Gateway &  & & & & & & & 5 \\
		&                                          &                                & Web Application Firewall & & & & & & & & 5 \\
		\hline
		Database             & Cloud SQL                          & SQL Database          & RDS & Trove & & & & & & & 5 \\
		&                                          & SQL Data Warehouse & Redshift  &  & & & & & & & 5 \\
		& Datastore                           & DocumentDB        & DynamoDB &  & & & St & & & & 5 \\
		& Big Table                            & Storage (Tables)  &  &  & & & & & & & 5 \\
		& Big Query                           &                            &  &  & & & & & & & 5 \\
		& Data Flow                           &                           & Data Pipeline &  & & & & & & & 5 \\
		& Data Proc                           & HDInsights            & EMR & Sahara & & & & & & & 5 \\
		&                                          & Data Lake Store   & & & & & & & & & 5 \\
		&                                          & Batch                   & &  & & & & & & & 5 \\     
		& Data Lab                            &                            & &  & & & & & & & 5 \\
		&                                          & Redis Cache          & ElastiCache & & & & & & & & 5 \\   
		&                                          &                            & Database Migration Service & & & & & & & & \textcircled{-} \\                                                    
		\hline
		Messaging           & PUB/SUB                            &                            &    &   & & & & & & & 5 \\
		&                                          & Notification Hubs  & SNS  & & & & & & & & 5 \\
		&                                          & Notification Hubs  & SES  &  & & & & & & & 5 \\                            
		&                                          & Event Hubs           & SQS + Lambda  & Zaqar & & & & & & & 5 \\                            
		&                                          & Storage (Queues) &   &  & & & & & & & 5 \\
		&                                          & Service Bus          &   &  & & & & & & & 5 \\                                           
		\hline
		Monitoring           & Cloud Monitoring                & Operational Insights & Cloud Watch & Ceilometer & & Mon & & & & & 4 \\
		& Cloud Logging                    &                              &  CloudTrail    & & & & & & & & 4 \\
		\hline
		Management        & Cloud Deployment Manager & Automation             & Cloud Formation   & Heat & & Sys & & & & Sys & 4 \\
		&                                          &                               & OpsWork &        &  &    & & & &   & 4 \\
		&                                          &                               & Config + Service Catalog &        &  &    & & & &   & \textcircled{-} \\
		&                                          & Azure Service Fabric &                         &  & & & & & & & 5 \\
		&                                          & Site Recovery          &                         &  & & & & & & & 5 \\
		&                                         & API Management      & API Gateway      &  & & & & & & & 5 \\    
		&                                         & Mobile Engagement  & Mobile Analytics  &  & & & & & & & 5 \\   
		&                                         & Active Directory  &  Directory Service  &  & & & & & & & 5 \\   
		&                                         & Multi Factor Authentication &  IAM               & Keystone & & & & & & & 4 \\   
		&                                         &                                          &  Certificate Manager & & & & & & & & 4 \\                                
		&                                         & Key Vault                           & CloudHSM           & Barbican & & & & & & & 4 \\                                
		&                                         & Security Center                   & Trusted Advisor  &  & & & & & & & \textcircled{-} \\
		&                                         &                                         & Inspector &  & & & & & & & \textcircled{-} \\
		
		\hline  
		Further Survices   & Translate API             &                            &                         &  & & & & & & & 5 \\
		& Prediction API           & Machine Learning     & Machine Learning &  & & & & & & & 5 \\    
		&                                & Data Lake Analytics  & Kinesis + QuickSight &  & & & & & & & 5 \\    
		&                                & Search                                 &   Cloud Search     &  & & & & & & & 5 \\    
		&                                & IoT Hub                                &   IoT                   &  & & & & & & & 5 \\    
		&                                & BizTalk Services                    &                           &  & & & & & & & 5 \\    
		&                                & VS Team Services  & CodeCommit       &  & & & & & & & \textcircled{-} \\    
		&                                & DevTests Labs                     &  CodePipeline\&Deploy & & & & & & & & \textcircled{-} \\    
		&                                & VS Application Insights &                           &  & & & & & & & 4 \\    
		&                                &  &  LumberJack                         &  & & & & & & & 5 \\    
		&                                &  &  WorkSpaces, Mail, Docs      &  & & & & & & & 6 \\                                 
		\hline
		\textbf{Sum of Services:}     & \textbf{22}              &  \textbf{46} & \textbf{55}   & \textbf{15}   & 5 & 6 & 4 & 1 & 2 & 1 & \\
		\hline
		\hline 
	\end{tabular}
	

	\smallskip
	\textbf{OCCI Concepts:} Compute = C, Network = N, Storage = St (used for CDMI as well)\\
	\textbf{CIMI Concepts:} System = Sys (used for TOSCA as well), Machine = M (used for OVF as well), Volume = V, Network = N, Monitoring = Mon\\
	\textbf{ClouNS Layers:} 1 = Physical, 2 = Virtual Node, 3 = Host, 4 = Cluster (Container/Storage), 5 = Service, 6 = Application
	
	\smallskip
	\textcircled{-} Currently not covered by ClouNS reference model. \\
	\textcircled{?} Standard is in a working state, unclear whether aspect will be covered.
	\vfill
\end{table*}

\noindent Table \ref{tab:cloud_services} shows the service portfolio of four major public cloud service providers and their relationship to current cloud standards. According to that, existing cloud standards cover only specific cloud service categories and do not show an integrated point of view. Instead of that, the proposed reference model covers almost all of these identified cloud services. Additionally, we discussed the suitability to describe a concrete cloud application in a structured way and checked conformance to the requirements for layered architectures and with our made experience derived from our action research activities. 

\textbf{The Social Collaboration Hub} is a SaaS offering that provides a complete intranet solution \cite{Peinl2015prowm} based on well-known open source products like \textit{Liferay} and \textit{Open Xchange} (OX). On the \textit{physical layer} \textbf{(Layer 1)} it builds on cheap commodity server with two CPUs / 32 cores, 128 GB RAM, 2 SSDs and 5x 1 TB hard disks each. They are connected with 10 GB ethernet. On the \textit{virtualization layer} \textbf{(Layer 2)}, \textit{OpenStack} with \textit{KVM} is used. \textit{Ceph} builds the storage cluster and provides block storage to \textit{OpenStack Cinder} using all hard disks and the SSDs as a cache. \textit{OpenStack Neutron} provides an SDN to connect the VMs \textbf{(Layer 2)}. Ubuntu 14.04 core LTS is used as operating system for the VMs \textbf{(Layer 3)} with an installed Docker runtime environment \textbf{(Layer 3 Container Host)}. VMs are dedicated to customers and not shared across them for better isolation and accountability. An \textit{Apache Mesos} cluster is providing an abstraction across all VMs \textbf{(Layer 4 Scheduler)}. Labels are used to assign Docker containers to the right VMs. \textit{Marathon} is used as an orchestrator and own extensions provide auto-scaling \textbf{(Layer 4 Orchestrator)}. Docker images are pulled from a private registry. Stateful containers write to a data partition mounted via \textit{Ceph} volume plug-in, so that storage can be easily migrated between hosts together with the container \textbf{(Layer 4 storage)}. \textit{Weave} is used as an overlay network to enable communication between containers across multiple host VMs \textbf{(Layer 4 Overlay Network)}. 

On \textbf{Layer 5}, a multitude of services is run. All data about user accounts is stored in an \textit{OpenLDAP} directory, other data is stored in a \textit{Percona XtraDB} cluster, which is based on \textit{MySQL} and \textit{Galera} cluster technology. Messaging is provided by \textit{Postfix} and \textit{Dovecot} for emails and an extension of \textit{Apache Shindig} for social communication. \textit{Shindig} runs its own \textit{Neo4j} database in embedded mode, which is in line with the microservice architecture that proposes freedom of choice for data stores. All these service provide only an API, but no user interface on their own. Real microservices with included GUI comprise \textit{CAS} for single sign-on, \textit{elasticsearch} for enterprise search, \textit{camunda} for workflows and business process management as well as \textit{Nuxeo} for document storage. 
The application itself is built on top of \textit{Liferay} for social collaboration and OX for groupware functionality \textbf{(Layer 6)}. 
\textit{Nuxeo} and \textit{Liferay} have own storage drivers for storing data in \textit{Amazon S3} compatible object stores, which are provided by \textit{Ceph}. From a categorization perspective, \textit{Liferay} and OX as well as the microservices do both have parts that belong to \textbf{Layer 6 (GUI)} and other parts that belong to \textbf{Layer 5 (API)}.

\textbf{Conformance with layered architectures and enterprise architecture methodologies}. Due to page limitations we can only provide some aspects of how we followed layered design principles. We started by a much more detailed layered architecture of 9 layers and reduced them to six layers in this proposal (Req 1). We reduced layers by shifting boundaries to reduce interactions between layers (Req 2). This layer reduction has been done intensively on Layer 4 but also on Layer 3. The reference model has layers for separate functions and technologies (Req. 3, 4). We cover IaaS, CaaS, PaaS and SaaS viewpoints for instance. Each layer has boundaries with its upper and lower layer only (Req. 5, 6 and 7). So concrete instances of a layer can be replaced (e.g. \textit{Kubernetes} could replace \textit{Mesos/Marathon} as a Layer 4 solution). Finally,  we cross checked and adapted these layers to be in accordance with EA frameworks like TOGAF. 

\section{Conclusion and outlook}
\label{sec:conclusion}
\noindent We use the presented ClouNS reference model quite successfully in our ongoing research to guide our technology identification, classification, adoption, research and development processes. Its main contribution for an enterprise architecture context is its guidance. For vendor lock-in aware enterprise architectures we advocate  to operate cloud-native applications on something that can be called a \textbf{portable cloud runtime environment} (Layer 4 according to our proposed reference model) to avoid vendor lock-in vulnerabilities coming along with higher cloud service layers. From the authors' point of view, the importance of these platforms for cloud computing is comparable to the  positive impact of the Java Virtual Machine specification on platform-independent application development in the 1995's.

Of course the reference model can be detailed in some aspects.  Especially Layer 5 services could be categorized more precisely to provide more guidance for enterprise architecture methodologies. This could be done by applying cloud service taxonomy systems \cite{HG2011}. For each layer and service category more detailed feature requirements and standardized data transferability requirements were desirable. This could be done by integrating requirement analysis outcomes presented for instance in \cite{PHP2016}. And finally, more detailed upward and downward standardization requirements for each layer were helpful to support vendor lock-in avoidance more systematically in enterprise architecture modeling \cite{PZF2013}. 
All this is up for ongoing research. However, the ClouNs reference model in its current state can already help to make
technologies, research approaches and standardization for cloud-native application development and enterprise architecture modeling more comparable, more codifyable and finally more integrated.

\subsubsection*{Acknowledgments}
\begin{small}
\noindent We would like to thank Dr. Adersberger from QAWARE GmbH. He inspired us with his thoughts about his MEKUNS approach (Mesos, Kubernetes, Cloud-native Stack). This research is funded by German Federal Ministry of Education and Research (Project Cloud TRANSIT, 03FH021PX4; Project SCHub, 03FH025PX4).
\end{small}



\bibliographystyle{IEEEtran}
\bibliography{IEEEabrv,kratzke}

\begin{thebibliography}{10}
\providecommand{\url}[1]{#1}
\csname url@samestyle\endcsname
\providecommand{\newblock}{\relax}
\providecommand{\bibinfo}[2]{#2}
\providecommand{\BIBentrySTDinterwordspacing}{\spaceskip=0pt\relax}
\providecommand{\BIBentryALTinterwordstretchfactor}{4}
\providecommand{\BIBentryALTinterwordspacing}{\spaceskip=\fontdimen2\font plus
\BIBentryALTinterwordstretchfactor\fontdimen3\font minus
  \fontdimen4\font\relax}
\providecommand{\BIBforeignlanguage}[2]{{%
\expandafter\ifx\csname l@#1\endcsname\relax
\typeout{** WARNING: IEEEtran.bst: No hyphenation pattern has been}%
\typeout{** loaded for the language `#1'. Using the pattern for}%
\typeout{** the default language instead.}%
\else
\language=\csname l@#1\endcsname
\fi
#2}}
\providecommand{\BIBdecl}{\relax}
\BIBdecl

\bibitem{PZF2013}
C.~Pahl, L.~Zhang, and F.~Fowley, ``{A Look at Cloud Architecture
  Interoperability through Standards},'' in \emph{{4th. Int. Conf. on Cloud
  Computing, GRIDs, and Virtualization (CLOUD COMPUTING 2013)}}, vol.~1, no.~1,
  2013, pp. 7--12.

\bibitem{CIMI}
\BIBentryALTinterwordspacing
M.~Hogan, L.~Fang, A.~Sokol, and J.~Tong, ``{Cloud Infrastructure Management
  Interface (CIMI) Model and RESTful HTTP-based Protocol, Version 2.0.0c},''
  2015. [Online]. Available:
  \url{http://www.nist.gov/customcf/get_pdf.cfm?pub_id=909024}
\BIBentrySTDinterwordspacing

\bibitem{OCCI-CORE}
\BIBentryALTinterwordspacing
R.~Nyren, A.~Edmonds, A.~Papaspyrou, and T.~Metsch, ``{Open Cloud Computing
  Interface (OCCI) - Core, Version 1.1},'' 2011. [Online]. Available:
  \url{https://www.ogf.org/documents/GFD.183.pdf}
\BIBentrySTDinterwordspacing

\bibitem{OCCI-IS}
\BIBentryALTinterwordspacing
T.~Metsch and A.~Edmonds, ``{Open Cloud Computing Interface (OCCI) -
  Infrastructure, Version 1.1},'' 2011. [Online]. Available:
  \url{https://www.ogf.org/documents/GFD.184.pdf}
\BIBentrySTDinterwordspacing

\bibitem{SHI+2013}
B.~Satzger, W.~Hummer, C.~Inzinger, P.~Leitner, and S.~Dustdar, ``{Winds of
  Change: From Vendor Lock-In to the Meta Cloud},'' \emph{{Internet Computing,
  IEEE}}, vol.~17, no.~1, pp. 69--73, Jan 2013.

\bibitem{FLR+2014}
C.~Fehling, F.~Leymann, R.~Retter, W.~Schupeck, and P.~Arbitter, \emph{Cloud
  Computing Patterns: Fundamentals to Design, Build, and Manage Cloud
  Applications}.\hskip 1em plus 0.5em minus 0.4em\relax Springer Publishing
  Company, Incorporated, 2014.

\bibitem{Stine2015}
M.~Stine, \emph{{Migrating to Cloud-Native Application Architectures}}.\hskip
  1em plus 0.5em minus 0.4em\relax O'Reilly, 2015.

\bibitem{BHJ2015}
A.~Balalaie, A.~Heydarnoori, and P.~Jamshidi, ``{Migrating to Cloud-Native
  Architectures Using Microservices: An Experience Report},'' in \emph{{1st
  Int. Workshop on Cloud Adoption and Migration (CloudWay)}}, Taormina, Italy,
  2015.

\bibitem{Newman2015}
S.~Newman, \emph{{Building Microservices}}.\hskip 1em plus 0.5em minus
  0.4em\relax {O'Reilly Media, Incorporated}, 2015.

\bibitem{namiot2014}
D.~Namiot and M.~Sneps-Sneppe, ``{On micro-services architecture},'' \emph{Int.
  Journal of Open Information Technologies}, vol.~2, no.~9, 2014.

\bibitem{Wiggins2012}
\BIBentryALTinterwordspacing
{Adam Wiggins}, ``{The Twelve-Factor App},'' 2014, last access 2016-02-14.
  [Online]. Available: \url{http://12factor.net/}
\BIBentrySTDinterwordspacing

\bibitem{Fowler2014}
\BIBentryALTinterwordspacing
{Martin Fowler}, ``{Circuit Breaker},'' 2014, last access 2016-05-27. [Online].
  Available: \url{http://martinfowler.com/bliki/CircuitBreaker.html}
\BIBentrySTDinterwordspacing

\bibitem{lockin2006}
\BIBentryALTinterwordspacing
{The Linux Information Project}, ``{Vendor Lock-in Definition},'' 2006, last
  access 2016-02-14. [Online]. Available:
  \url{http://www.linfo.org/vendor_lockin.html}
\BIBentrySTDinterwordspacing

\bibitem{Arthur1989}
W.~B. Arthur, ``{Competing Technologies, Increasing Returns, and Lock-In by
  Historical Events},'' \emph{{The Economic Journal}}, vol.~99, no. 394, pp.
  116--131, 1989.

\bibitem{di2015class}
B.~Di~Martino, G.~Cretella, and A.~Esposito, ``{Classification and Positioning
  of Cloud Definitions and Use Case Scenarios for Portability and
  Interoperability},'' in \emph{{3rd Int. Conf. on Future Internet of Things
  and Cloud (FiCloud 2015)}}.\hskip 1em plus 0.5em minus 0.4em\relax IEEE,
  2015, pp. 538--544.

\bibitem{KQ2015}
N.~Kratzke and P.-C. Quint, ``{About Automatic Benchmarking of IaaS Cloud
  Service Providers for a World of Container Clusters},'' \emph{{Journal of
  Cloud Computing Research}}, vol.~1, no.~1, pp. 16--34, 2015.

\bibitem{Kra2014a}
N.~Kratzke, ``{Lightweight Virtualization Cluster - How to overcome Cloud
  Vendor Lock-In},'' \emph{{Journal of Computer and Communication (JCC)}},
  vol.~2, no.~12, Oct. 2014.

\bibitem{OSF2014}
J.~Opara-Martins, R.~Sahandi, and F.~Tian, ``Critical review of vendor lock-in
  and its impact on adoption of cloud computing,'' in \emph{Int. Conf. on
  Information Society (i-Society 2014)}, Nov 2014, pp. 92--97.

\bibitem{HFST2011}
\BIBentryALTinterwordspacing
J.~Durand, M.~Andreou, D.~Davis, and G.~Pilz, ``{NIST Cloud Computing Standards
  Roadmap},'' 2011. [Online]. Available:
  \url{https://www.dmtf.org/sites/default/files/standards/documents/DSP0263_2.0.0c.pdf}
\BIBentrySTDinterwordspacing

\bibitem{OVF}
\BIBentryALTinterwordspacing
{System Virtualization, Partitioning, and Clustering Working Group}, ``{Open
  Virtualization Format Specification, Version 2.1.0},'' 2015. [Online].
  Available:
  \url{https://www.dmtf.org/sites/default/files/standards/documents/DSP0243_2.1.1.pdf}
\BIBentrySTDinterwordspacing

\bibitem{CDMI}
\BIBentryALTinterwordspacing
SNIA, ``{Cloud Data Management Interface (CDMI), Version 1.1},'' 2015.
  [Online]. Available:
  \url{http://www.snia.org/sites/default/files/CDMI_Spec_v1.1.1.pdf}
\BIBentrySTDinterwordspacing

\bibitem{SIIF}
\BIBentryALTinterwordspacing
{Intercloud WG (ICWG) Working Group}, ``{Standard for Intercloud
  Interoperability and Federation (SIIF)}.'' [Online]. Available:
  \url{https://standards.ieee.org/develop/project/2302.html}
\BIBentrySTDinterwordspacing

\bibitem{CPIP}
\BIBentryALTinterwordspacing
------, ``{Guide for Cloud Portability and Interoperability Profiles (CPIP)}.''
  [Online]. Available:
  \url{https://standards.ieee.org/develop/project/2301.html}
\BIBentrySTDinterwordspacing

\bibitem{dimartino2015}
B.~Di~Martino, G.~Cretella, and A.~Esposito, ``{Cross-platform cloud APIs},''
  in \emph{{Cloud Portability and Interoperability}}.\hskip 1em plus 0.5em
  minus 0.4em\relax Springer, 2015, pp. 45--57.

\bibitem{JD2011}
G.~Juve and E.~Deelman, ``Automating application deployment in infrastructure
  clouds,'' in \emph{{3rd Int. Conf. on Cloud Computing Technology and Science
  (CLOUDCOM 2011)}}.\hskip 1em plus 0.5em minus 0.4em\relax Washington, DC,
  USA: IEEE Computer Society, 2011, pp. 658--665.

\bibitem{ACA+2011}
C.~de~Alfonso, M.~Caballer, F.~Alvarruiz, G.~Molto, and V.~Hernandez,
  ``{Infrastructure Deployment Over the Cloud},'' in \emph{{3rd Int. Conf. on
  Cloud Computing Technology and Science (CLOUDCOM 2011)}}, Nov 2011, pp.
  517--521.

\bibitem{LDK+2011}
A.~Lenk, C.~Danschel, M.~Klems, D.~Bermbach, and T.~Kurze, ``{Requirements for
  an IaaS deployment language in federated Clouds},'' in \emph{{Int. Conf. on
  Service-Oriented Computing and Applications (SOCA 2011)}}, Dec 2011, pp.
  1--4.

\bibitem{MGGD2012}
S.~Murphy, S.~Gallant, C.~Gaughan, and M.~Diego, ``{U.S. Army Modeling and
  Simulation Executable Architecture Deployment Cloud Virtualization
  Strategy},'' in \emph{{12th Int. Symp. on Cluster, Cloud and Grid Computing
  (CCGrid)}}, May 2012, pp. 880--885.

\bibitem{TOSCA}
\BIBentryALTinterwordspacing
{OASIS}, ``{Topology and Orchestration Specification for Cloud Applications
  (TOSCA), Version 1.0},'' 2013. [Online]. Available:
  \url{http://docs.oasis-open.org/tosca/TOSCA/v1.0/os/TOSCA-v1.0-os.pdf}
\BIBentrySTDinterwordspacing

\bibitem{LTC+2014}
D.-H. Le, H.-L. Truong, G.~Copil, S.~Nastic, and S.~Dustdar, ``{SALSA: A
  Framework for Dynamic Configuration of Cloud Services},'' in \emph{6th Int.
  Conf. on Cloud Computing Technology and Science (CloudCom 2014)}, Dec 2014,
  pp. 146--153.

\bibitem{zlsd2015}
R.~Zabolotnyi, P.~Leitner, S.~Schulte, and S.~Dustdar, ``{SPEEDL - A
  Declarative Event-Based Language for Cloud Scaling Definition},'' in
  \emph{{IEEE World Congress on Services (SERVICES 2015)}}, 2015.

\bibitem{CMT+2013}
G.~Copil, D.~Moldovan, H.~L. Truong, and S.~Dustdar, ``{SYBL: An Extensible
  Language for Controlling Elasticity in Cloud Applications},'' in \emph{{13th
  IEEE/ACM Int. Symp. on Cluster, Cloud and Grid Computing (CCGrid
  2013)}}.\hskip 1em plus 0.5em minus 0.4em\relax IEEE Computer Society, 2013,
  pp. 112--119.

\bibitem{ISO-7498}
\BIBentryALTinterwordspacing
{ISO/OSI}, ``{Open Systems Interconnection - Basic Reference Model: The Basic
  Model},'' 1994. [Online]. Available:
  \url{http://standards.iso.org/ittf/PubliclyAvailableStandards/s020269_ISO_IEC_7498-1_1994(E).zip}
\BIBentrySTDinterwordspacing

\bibitem{WBM+2010}
K.~Winter, S.~Buckl, F.~Matthes, and C.~M. Schweda, ``Investigating the
  state-of-the-art in enterprise architecture management methods in literature
  and practice.'' \emph{Mediterranean Conference on Information Systems (MCIS
  2010)}, vol.~90, 2010.

\bibitem{Zachman2008}
\BIBentryALTinterwordspacing
J.~Zachman, ``{John Zachman's Concise Definition of the Zachman Framework},''
  2008. [Online]. Available:
  \url{https://www.zachman.com/about-the-zachman-framework}
\BIBentrySTDinterwordspacing

\bibitem{MG2011}
P.~M. Mell and T.~Grance, ``{The NIST Definition of Cloud Computing},''
  {National Institute of Standards \& Technology}, Gaithersburg, MD, United
  States, Tech. Rep., 2011.

\bibitem{BML+2011}
R.~B. Bohn, J.~Messina, F.~Liu, J.~Tong, and J.~Mao, ``Nist cloud computing
  reference architecture,'' in \emph{{World Congr. on Services (SERVICES
  2011)}}.\hskip 1em plus 0.5em minus 0.4em\relax Washington, DC, USA: IEEE
  Computer Society, 2011, pp. 594--596.

\bibitem{POSIX}
\BIBentryALTinterwordspacing
IEEE and OpenGroup, ``{The Open Group Base Specifications Issue 7, IEEE Std
  1003.1, 2013 Edition (POSIX Standard)},'' 2013. [Online]. Available:
  \url{http://pubs.opengroup.org/onlinepubs/9699919799/}
\BIBentrySTDinterwordspacing

\bibitem{PH2015}
R.~Peinl and F.~Holzschuher, ``{The Docker Ecosystem Needs Consolidation},'' in
  \emph{{5th Int. Conf. on Cloud Computing and Services Science (CLOSER
  2015)}}, 2015, pp. 535--542.

\bibitem{OCI}
\BIBentryALTinterwordspacing
OCI, ``{Open Container Initiative},'' 2015, last access 2016-02-04. [Online].
  Available: \url{https://www.opencontainers.org}
\BIBentrySTDinterwordspacing

\bibitem{caas}
\BIBentryALTinterwordspacing
{Docker Inc.}, ``{Containers as a Service (CaaS) as your new platform for
  application development and operations},'' 2016, latest access 2016-02-05.
  [Online]. Available:
  \url{https://blog.docker.com/2016/02/containers-as-a-service-caas/}
\BIBentrySTDinterwordspacing

\bibitem{mesos}
B.~Hindman, A.~Konwinski, M.~Zaharia, A.~Ghodsi, A.~D. Joseph, R.~H. Katz,
  S.~Shenker, and I.~Stoica, ``{Mesos: A Platform for Fine-Grained Resource
  Sharing in the Data Center.}'' in \emph{{8th USENIX Conf. on Networked
  systems design and implementation (NSDI'11)}}, vol.~11, 2011.

\bibitem{kubernetes}
A.~Verma, L.~Pedrosa, M.~R. Korupolu, D.~Oppenheimer, E.~Tune, and J.~Wilkes,
  ``{Large-scale cluster management at Google with Borg},'' in \emph{{10th.
  Europ. Conf. on Computer Systems (EuroSys '15)}}, Bordeaux, France, 2015.

\bibitem{QK2016a}
P.-C. Quint and N.~Kratzke, ``{Overcome Vendor Lock-In by Integrating Already
  Available Container Technologies - Towards Transferability in Cloud Computing
  for SMEs},'' in \emph{{7th. Int. Conf. on Cloud Computing, GRIDS and
  Virtualization (CLOUD COMPUTING 2016)}}, 2016.

\bibitem{weave}
\BIBentryALTinterwordspacing
Weave, ``{Weave Net},'' 2016, last access 2016-02-05. [Online]. Available:
  \url{http://weave.works/products/weave-net/}
\BIBentrySTDinterwordspacing

\bibitem{calico}
\BIBentryALTinterwordspacing
Calico, ``{Calico - A Pure Layer 3 Approach to Virtual Networking for Highly
  Scalable Data Centers},'' 2016, last access 2016-02-05. [Online]. Available:
  \url{http://www.projectcalico.org/}
\BIBentrySTDinterwordspacing

\bibitem{flannel}
\BIBentryALTinterwordspacing
CoreOS, ``{flannel - a virtual network that gives a subnet to each host for use
  with container runtimes},'' 2016, last access 2016-02-05. [Online].
  Available: \url{https://github.com/coreos/flannel}
\BIBentrySTDinterwordspacing

\bibitem{PHP2016}
R.~Peinl, F.~Holzschuher, and F.~Pfizer, ``{Docker cluster management - survey
  results and future},'' \emph{{Journal of Grid Computing}}, 2016, to be
  published.

\bibitem{Peinl2015prowm}
R.~Peinl, ``{Supporting Knowledge Management Instruments with Composable
  Micro-Services},'' in \emph{{Wissensgemeinschaften - ProWM 2015, Dresden,
  Juni 2015}}, 2015.

\bibitem{HG2011}
C.~N. H{\"o}fer and G.~Karagiannis, ``Cloud computing services: taxonomy and
  comparison,'' \emph{{Journal of Internet Services and Applications}}, vol.~2,
  no.~2, pp. 81--94, 2011.

\end{thebibliography}
%


\end{document}